\documentclass{iaus}
\usepackage{graphicx}

\title[Binary Evolutionary Models] 
{Binary Evolutionary Models}

\author[Han \& Podsiadlowski]   
{Z. Han$^1$ \& Ph. Podsiadlowski$^2$}

\affiliation{
$^1$National Astronomical Observatories /
 Yunnan Observatory, Kunming, 65011, P.R. China
\break email: zhanwenhan@hotmail.com\\[\affilskip]
$^2$Department of Physics, University of Oxford, Keble Road, 
Oxford OX1 3RH, UK
\break email: podsi@astro.ox.ac.uk}

\pubyear{2008}
\volume{252}  
\pagerange{119--126}
\date{?? and in revised form ??}
\jname{The Art of Modelling Stars in the 21st Century}
\editors{L. Deng, K.L. Chan \& C. Chiosi, eds.}
\begin{document}

\maketitle

\begin{abstract}
In this talk, we present the general principles of binary evolution 
and give two examples. The first example is the formation
of subdwarf B stars (sdBs) and their application to the long-standing
problem of ultraviolet excess (also known as UV-upturn) in elliptical 
galaxies. The second is for the progenitors of type Ia supernovae
(SNe Ia). 
We discuss the main binary interactions, i.e.,
stable Roche lobe overflow (RLOF) and common envelope (CE) evolution, 
and show evolutionary channels 
leading to the formation of various binary-related objects.
In the first example,
we show that the binary model of sdB stars of Han et al.
(2002, 2003) can reproduce field sdB stars and their counterparts,
extreme horizontal branch (EHB) stars, in globular clusters.
By applying the binary model to the study of evolutionary population
synthesis, we have obtained an ``a priori'' model for the UV-upturn
of elliptical galaxies and showed that the UV-upturn is most likely
resulted from binary interactions. This has major implications for 
understanding the evolution of the UV excess and elliptical galaxies
in general.
In the second example, we introduce the single degenerate channel
and the double degenerate channel for the progenitors of SNe Ia.
We give the birth rates and delay time distributions for each channel
and the distributions of companion stars at the moment of SN explosion
for the single degenerate channel, which would help to search for
the remnant companion stars observationally.
\end{abstract}

\firstsection 
\section{General Principles of Binary Evolution}

About half of the stars are in binaries and binary evolution 
plays a crucial role in the formation of 
many interesting objects, such as Algols, FK Comae (FK Com) stars, cataclysmic
variables (CVs), planetary nebulae (PNe), barium (Ba) stars, CH stars, 
type Ia supernovae (SNe Ia), AM Canum Venaticorum (AM CVn) stars, 
low mass X-ray binaries (LMXB), high mass X-ray binaries (HMXB),
symbiotic (Sym) stars, blue stragglers (BSs), pulsars, subdwarf B
(sdB) stars, double degenerates (DDs) etc. Binary evolution is also 
important in the study of evolutionary population synthesis,
which is a power tool in the study of galaxies.

A binary system (of low/intermediate mass) has two components: 
the primary (the initially more massive
one) and the secondary. As the binary evolves, the primary expands and 
may fill its Roche lobe on the Hertzsprung gap or on the giant branch.
Roche lobe overflow (RLOF) begins and 
the primary's envelope mass transfers to the secondary.
Given the mass 
ratio $q$ of primary to secondary less than a critical value $q_{\rm c}$ 
(\cite[Hjellming \& Webbink, 1987]{hje87}; \cite[Webbink, 1988]{hje88};
\cite[Han \& Webbink, 1999]{han99}; \cite[Han et al., 2002]{han02})
at the onset of the mass transfer, 
where $q_{\rm c}$ mainly depends on the entropy profile
of the primary's envelope and the angular momentum loss from the system, 
the mass transfer is stable, leading to a wide white dwarf (WD) binary. 
Given the mass ratio $q$ larger than $q_{\rm c}$, the mass transfer is
dynamically unstable, leading to the formation of a common envelope
(CE; \cite[Paczy\'nski, 1976]{pac76}). 
The CE engulfs the core of the primary and the secondary, and
does not co-rotate with the embedded binary. The friction between the CE
and the embedded binary makes the orbit decay, and a large amount of
orbital energy released is deposited into the CE. If the CE can be ejected,
a close WD binary forms, otherwise a fast rotating merger is resulted.
For a WD binary system, the secondary continues to evolve and may 
experience mass transfer. Similar to the process described above,
the mass transfer may lead to the formation of a CE and the CE ejection
produces a double degenerate.

\begin{figure}
\centerline{
\includegraphics[width=9.0cm]{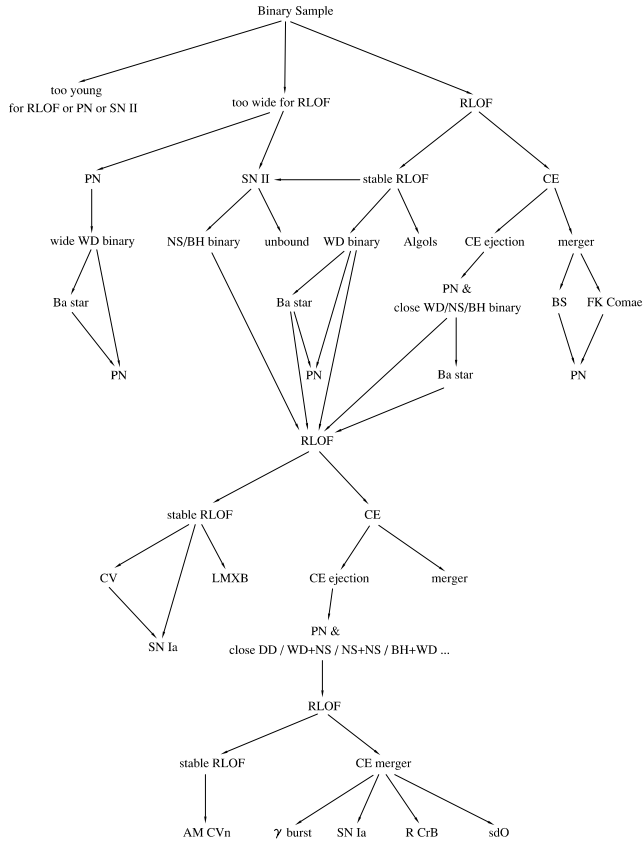}
}
\caption{
A simplified version of flow chart of binary evolution. See 
paragraph 1 of Section 1 of the text for the acronyms.
}
\label{flow}
\end{figure}

Figure~\ref{flow} is a flow chart of binary evolution. It is by
no means comprehensive, but it does show evolutionary channels
leading to various objects for a binary system with given conditions.
 
A good stellar evolution theory should give and predict the statistical
properties of a stellar population as well as the properties of 
individual stars or binaries. Binary population synthesis (BPS) is to 
evolve a large number of stars (including binaries) in order to
investigate statistical properties of stars and check evolutionary
mechanisms for different types of stars. In a BPS study, we first
generate a binary sample (10 million binaries), then evolve the sample
according to stellar evolution model grids and taking into account
binary interactions, and we obtain different types of binary-related
objects, which can be directly compared to observations.

In addition to the grids of stellar evolutionary models,
we adopt the following input in BPS simulations: 
\begin{description}
\item (1) A constant star-formation rate is taken  
          over the last 15\,Gyr for a field population, or
          alternatively, a single star burst for a globular cluster
          or an elliptical galaxy.
\item (2) The initial mass function of \cite{mil79} 
          is adopted. 
\item (3) We mainly adopt a constant initial mass ratio distribution
          $n(q')$, where $q'=1/q$ is the ratio of secondary to primary. 
\item (4) We take the distribution of
    separations to be constant in $\log a$ for wide binaries, where $a$ is
    the orbital separation.  The adopted distribution gives that $\sim
    50$\,\% of stellar systems are binary systems with orbital periods
    less than 100 yr.
\end{description}

The main model parameters in BPS are 
mass transfer efficiency  $\alpha_{\rm RLOF}$ for the first stable RLOF
(where the accretor is a main sequence star), CE ejection efficiency 
$\alpha_{\rm CE}$ and thermal contribution $\alpha_{\rm th}$ for
CE evolution. 
The mass transfer efficiency  $\alpha_{\rm RLOF}$ 
is the fraction of the envelope mass that is transferred
onto the secondary rather than is ejected from the system, 
where we assume that the matter lost from the system carries away
the specific angular momentum of the system. A typical value
for $\alpha_{\rm RLOF}$ is 0.5.
The CE ejection efficiency 
$\alpha_{\rm CE}$ is the fraction of the released 
orbital energy used to overcome the binding energy of the envelope
during the spiral-in process of a CE.
The thermal contribution $\alpha_{\rm th}$ defines
the fraction of the internal energy of thermodynamics (including
recombination energy as well as the thermal energy) 
contributing to the binding energy of the CE. A CE is ejected if
\begin{equation}
\alpha_{\rm CE}\Delta E_{\rm orb} 
\ge E_{\rm gr} - \alpha_{\rm th} E_{\rm th}\, ,
\end{equation}
where $\Delta E_{\rm orb}$ is the orbital energy released during the
spiral-in process, $E_{\rm gr}$ the gravitational binding energy of the
CE, $E_{\rm th}$ the internal energy of the CE. Both $E_{\rm gr}$ and
$E_{\rm th}$ are calculated from detailed stellar models and therefore
the prescription here is different from the $\lambda$ prescription
but appear to be more physical. We refer the reader to 
\cite{han94}, \cite{dew00} and \cite{pod03} for the details.
The inclusion of the internal energy in ejecting the CE seems to be
the most plausible in the explanation of both long-period and short-period
binaries containing compact objects 
(\cite[Han et al., 1995a; 1995b]{han95a,han95b};
\cite[Dewi \& Tauris, 2000]{dew00}). 
\cite{web07} has made a physical
investigation on both the energetics of CE evolution and 
the angular momentum prescription (i.e. the $\gamma$ prescription,
\cite[Nelemans et al., 2000]{nel00}), and convincingly showed 
the necessity of recombination energy term for common
envelope evolution. Previous studies, e.g., 
\cite{han95a}, \cite{han95b}, \cite{han98}, \cite{han02},
\cite{han03}, \cite{han04}, have showed that both $\alpha_{\rm th}$
and $\alpha_{\rm th}$ are close to one.

\section{The binary model for subdwarf B stars and the UV-upturn 
of elliptical galaxies}

Subdwarf B (sdB) stars
\footnote{In this paper, we collectively refer to helium-core-burning
stars with thin hydrogen envelopes as sdB stars, even if some of
them may in reality be sdO or sdOB stars}
are core helium-burning stars with very thin
hydrogen envelope (\cite[Heber, 1986]{heb96}).
They are important in many aspects of astrophysics,
e.g., stellar evolution, distance indicators, Galactic structure,
and the long-standing problem of far-ultraviolet excess 
in early-type galaxies 
(\cite[Kilkenny et al., 1997]{kil97}; 
 \cite[Green, Schmidt \& Liebert, 1986]{gre86};
 \cite[Han, Podsiadlowski \& Lynas-Gray, 2007]{han07}).

\cite{max01} showed that the majority of field sdB stars are in
binaries, and this has posed a serious challenge to stellar
evolution theory. Han et al.\ (\cite[2002; 2003]{han02,han03})
proposed a binary model for their formation. 
In the model, there are three types of formation channels 
for sdB stars: stable RLOF for sdB binaries with long orbital periods,
CE ejection for sdB binaries with short orbital periods, and the merger
of helium WDs to form single EHB stars.
In the stable RLOF channel, the mass donor fills its Roche lobe
near the tip of the first giant branch and experiences a stable
mass transfer, and its envelope is striped off by the RLOF, 
and the naked helium core (with thin hydrogen envelope)
get ignited to produce a sdB stars.
In the CE ejection channel, the mass donor also fills its Roche lobe
near the tip of the first giant branch
to have a dynamically unstable mass transfer leading to the formation of
a CE. The CE ejection leaves a naked helium core (with thin hydrogen
envelope) and the naked helium core is ignited to produce a sdB star.
In the CE channel, the donor star needs to fill its Roche lobe
closer to the tip of the first giant branch, or, in other words,
the minimum core mass required for the donor star at the onset
of mass transfer is larger than that in the stable RLOF channel to 
produce a sdB star. This is simply because that the time scale of 
the CE evolution is much shorter than that of the stable RLOF and the core
does not grow by much in the CE process.
In the merger channel, a close helium WD pair coalesces due to
angular momentum loss via gravitational wave radiation.
The binary model of Han et al. (\cite[2002; 2003]{han02,han03})
has successfully explained the main observational 
characteristics of field sdB stars:
their distributions in the orbital period-minimum
companion mass diagram, and in the effective temperature-surface
gravity diagram; their distributions of orbital period and mass
function; their binary fraction and the fraction of sdB
binaries with WD companions; their birth rates; and their
space density. The model is indeed a step forward
and is widely used in the study of sdB stars 
(\cite[O'Tool, Heber \& Benjamin, 2004]{oto04}).

\cite{mon06,mon08}
have done radial-velocity surveys for extreme horizontal branch (EHB)
stars, the counterparts of field sdB stars, in globular clusters.
They found that there is a remarkable lack of close 
binary systems in EHB stars. This is surprising as compared to the
high binary fraction in field sdB stars. They speculated that
there may exist a binary fraction-age relation for sdB stars.
\cite{han08b} showed that such a relation does exist and 
the binary model of Han et al.
(\cite[2002; 2003]{han02,han03}) can reproduce the EHB stars in
globular clusters, in particular, the low binary fraction of the 
EHB stars. The main reason for the low binary fraction is that
the stars in a globular cluster are all old, and the envelopes
of donor stars in the CE channel are loosely bound, leading to
wide EHB binaries rather than close ones.

One of the first major discoveries soon after the advent of UV
astronomy was the discovery of an excess of light in the
far-ultraviolet (far-UV) in elliptical galaxies (see the review by
O'Connell, \cite[1999]{oco99}).
This came as a complete surprise since
elliptical galaxies were supposed to be entirely composed of old, red
stars and not to contain any young stars that radiate in the UV.
Since then it has become clear that the far-UV excess (or upturn) is
not a sign of active contemporary star formation, but is caused by an
older population of helium-burning stars or their descendants with a
characteristic surface temperature of 25,000\,K 
(\cite[Ferguson et al., 1991]{fer91}).

The origin of this population of hot, blue stars in an otherwise red
population has, however, remained a major mystery.
As we described above, the binary model of Han et al.\ 
(\cite[2002; 2003]{han02,han03}) reproduces Galactic hot subdwarfs
(synonymous with sdB stars in this paper). 
The key feature of the channels in the model is that 
they provide the missing physical mechanism
for ejecting the envelope and for producing a hot subdwarf. 
Moreover,
since it is known that these hot subdwarfs provide an important source
of far-UV light in our own Galaxy, it is not only reasonable to assume
that they will also contribute significantly to the far-UV in
elliptical galaxies, but is in fact expected.
It would, therefore, be ``a priori'' to apply the Han et al. model
to the study of the UV-upturn problem.

To quantify the importance of the effects of binary interactions
on the spectral appearance of elliptical galaxies, we have
performed the first population synthesis
study of galaxies that includes binary evolution 
(see also \cite[Bruzual \& Charlot, 1994]{bru94};
\cite[Worthy, 1994]{wor94}; \cite[Zhang, Li \& Han, 2005]{zha05}).
It is based on a binary population synthesis model of Han et al.
(\cite[2002; 2003]{han02,han03})
that has been calibrated to reproduce the short-period
hot subdwarf binaries in our own Galaxy that make up the majority of
Galactic hot subdwarfs (\cite[Maxted et al., 2001]{max01}). 
The population synthesis model
follows the detailed time evolution of both single and binary stars,
including all binary interactions, and is capable of simulating
galaxies of arbitrary complexity, provided that the star-formation history
is specified. To obtain galaxy colours and spectra, we have calculated
detailed grids of spectra for hot subdwarfs using the {\scriptsize
ATLAS9} (\cite[Kurucz, 1992]{kur92}) stellar atmosphere code,
which calculates
plane-parallel atmospheres in local thermodynamic equilibrium.
For the spectra and colours of
single stars with hydrogen-rich envelopes, we use the comprehensive
BaSeL library of theoretical stellar spectra 
(\cite[Lejeune, Cuisinier \& Buser, 1997; 1998]{lej97,lej98}) 

\begin{figure}
\centerline{
\includegraphics[height=9.0cm,angle=270]{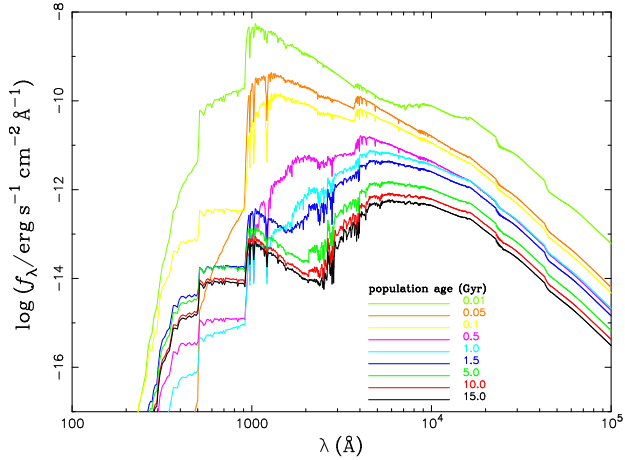}
}
\caption{
The evolution of the rest-frame intrinsic spectral
energy distribution (SED) for a simulated galaxy
in which all stars formed at the same time, i.e. a simple stellar population
(SSP).  The stellar population (including
binaries) has a mass of $10^{11}M_\odot$ and the galaxy is assumed to
be at a distance of 10\,Mpc.  The figure is for
the standard simulation set (with $\alpha_{\rm CE}=\alpha_{\rm th}=0.75$
and $\alpha_{\rm RLOF}=0.5$)
in \cite{han07}
and no offset is applied to the SEDs. Note that the 
line sections 
of between 500\AA\ and 900\AA \  for population ages of 1.5 and 5.0\,Gyr
overlap. 
}
\label{sed}
\end{figure}

Figure~\ref{sed} shows our simulated evolution of the spectral energy
distribution (SED) of a galaxy in which all the stars formed at the
same time.  The total mass of the stellar population (including
binaries) is $10^{11}M_\odot$ and the galaxy is taken to be at a
distance of 10\,Mpc.  At early times, the far-UV flux is entirely
caused by the contribution from young stars. Hot subdwarfs from the
various binary evolution channels become important after about
1.1\,Gyr, which corresponds to the evolutionary timescale of a
2\,$M_{\odot}$ star, and soon start to dominate completely. After a
few Gyr the far-UV SED no longer changes appreciably relative to the
visual flux. One immediate implication of this is that the model
predicts that the magnitude of the UV excess $(1550-V)$, defined as
the relative ratio of the flux in the $V$ band to the far-UV flux
(\cite[Burstein et al., 1988]{bur88}), 
should not evolve significantly with look-back time or
redshift. Indeed, this is exactly what seems to have been found in
recent observations 
(\cite[Brown et al., 2003; Rich et al., 2005]{bro03,ric05}).

We found that our binary model can naturally explain many observations
of early-type galaxies in spite of its simplicity, 
and UV-upturn is expected to be universal (from
dwarf to giant ellipticals; see \cite[Lisker \& Han, 2008]{lis08}). 
The model also predicts that
the magnitude of UV-upturn does not depend much on metallicity or 
redshift.
We refer the reader to \cite{han07} for the details.

\section{Progenitors of Type Ia supernovae}

Recent progress in cosmology is largely due to the use of
Type Ia supernovae (SNe Ia) as a {\it calibrated} distance indicator
(\cite[Riess et al., 1998; Perlmutter et al., 1999]{rie98,per99}).
 The nature of their progenitors is still unclear, 
raising doubts as to the calibration which is purely empirical and based
on nearby SN Ia sample. The SNe Ia are believed to be
thermonuclear explosions of carbon-oxygen (CO) WDs. 
Observational characteristics of SNe Ia imply that
the explosion occurs when a CO WD reaches the Chandrasekhar limit.
There are mainly two channels to create Chandrasekhar-mass CO WDs:
{\it the single degenerate channel}, where the CO WD accretes mass from 
a non-degenerate companion (\cite[Hachisu, Kato \& Nomoto, 1999a; 
Han \& Podsiadlowski, 2004]{hac99a,han04}), and 
{\it the double degenerate channel}, where two CO WDs with a 
total mass larger than the Chandrasekhar mass coalesce 
(\cite[Iben \& Tutukov, 1984; Webbink \& Iben, 1987]{ibe84,web87})
\footnote{Note, however, that in this case it is quite likely that
the merger product experiences core collapse rather than
a thermonuclear explosion(\cite[Nomoto \& Iben, 1985]{nom85}).}.     

\begin{figure}
\centerline{
\includegraphics[height=9.0cm,angle=270]{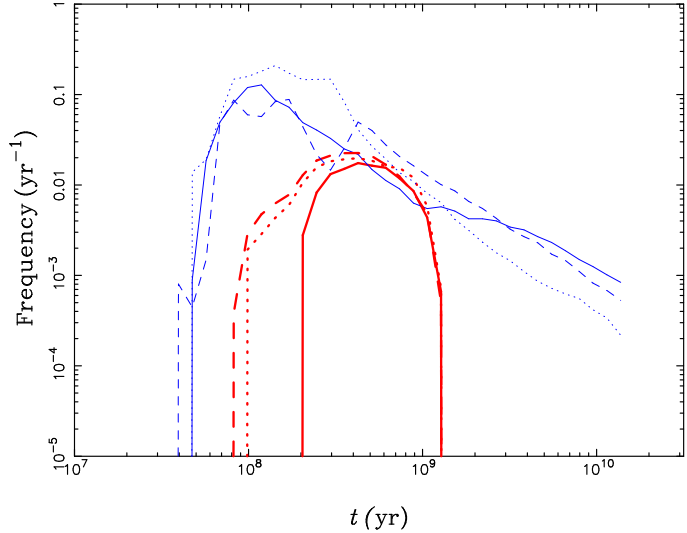}
}
\caption{
The evolution of 
birthrates of SNe Ia for a single star burst of $10^{11}M_\odot$
of solar metallicity. Solid, dashed and dotted lines are
for $\alpha_{\rm CE}=\alpha_{\rm th}=1.0$, 0.75, 0.5, respectively
($\alpha_{\rm RLOF}=0.5$). Thin lines are for the double degenerate
channel, and thick lines are for the single degenerate channel.
}
\label{rate}
\end{figure}

Employing Eggelton's stellar evolution code 
(\cite[Eggleton, 1971; 1972; 1973; Han, Podsiadlowski \& Eggleton, 1994;
Pols et al., 1995]{egg71,egg72,egg73,han94,pol95}) 
and adopting the prescription of \cite{hac99b} for the accretion
efficiency of a CO WD, 
\cite{han04} carried out detailed binary evolution calculations
for about 2300 close CO WD binaries, and mapped out the initial 
parameters in the orbital period-secondary mass plane (for a range
of WD masses) which lead to a SN Ia. They have implemented
these results in a binary population synthesis (BPS) study to obtain
the birth rates for SNe Ia for a constant star formation rate.
The Galactic birth rate is lower than (but comparable to) that
inferred observationally.
They have also obtained the evolution of birth rates 
with time for a single star burst.
We see from Fig.~\ref{rate} that the time delay of SN Ia explosion
from star burst is $\sim 0.1$ to $\sim 1$\,Gyr for the single
degenerate channel.
The birth rates from the double degenerate channel reach to peaks
at $\sim 0.1$\,Gyr and decays with age (approximately $\propto t^{-1}$).
\cite{men08} did similar investigations, but for 10 metallicities
in order to investigate the metallicity effect.

\begin{figure}
\centerline{
\includegraphics[height=9.0cm,angle=270]{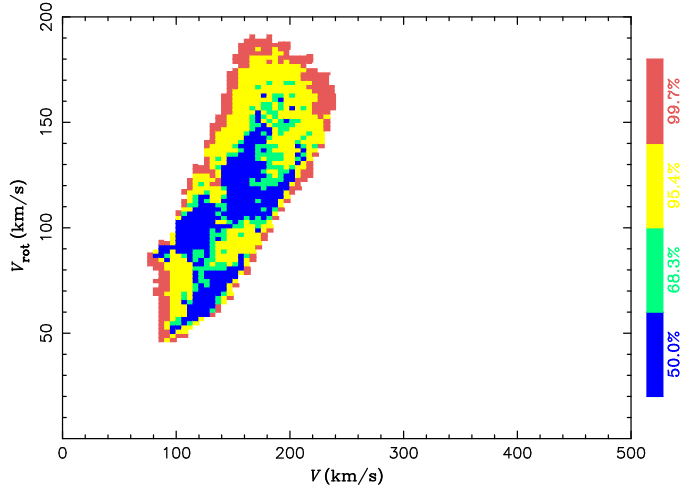}
}
\caption{
Snapshot probability distribution of companion stars in the
plane of ($V_{\rm orb}$, $V_{\rm rot}$) at the current 
epoch, where $V_{\rm orb}$ is the orbital velocity and $V_{\rm rot}$
the equatorial rotational velocity for the companion stars
at the moment of SN explosion. The probability decreases from
inner regions to outer regions.
Regions, from inside to outside with corresponding
gradational grey scale in the legend (from bottom to top), together 
with the inner regions, contain 50.0\%,
68.3\%, 95.4\%, and 99.7\% of all the systems, respectively.
The model adopts a constant star formation rate over the
last 15\,Gyr and $\alpha_{\rm CE}=\alpha_{\rm th}=0.75$
($\alpha_{\rm RLOF}=0.5$).
For a similar model but with
$\alpha_{\rm CE}=\alpha_{\rm th}=1.0$, the distribution is
similar, but the upper edge of 190\,${\rm km/s}$ moves down to
170\,${\rm km/s}$.
}
\label{vel}
\end{figure}

For the single degenerate model, the remnant companion star
after SN explosion would be a fast
rotator and have a high space velocity. Fig.~\ref{vel} is the distribution
of companion star in the plane of orbital velocity-rotational velocity
at the moment of SN explosion, as derived from a BPS study with
the implementation of the results of \cite{han04}. Note, however, the
ejecta of SN explosion would impact the companion and the companion
obtains a kick velocity and the total velocity may be higher by up to
10\% (\cite[Meng, Chen \& Han, 2007]{men07}). See \cite{han08a} for
more distributions of companion stars.

\begin{acknowledgments}
This work was in part supported by the Natural Science Foundation of China 
under Grant Nos 10433030, 10521001 and 2007CB815406.
\end{acknowledgments}

\end{document}